\documentclass[usegraphicx,useAMS,usenatbib]{mn2e}
\usepackage{times}

\newcommand{\apj}{ApJ}

\newcommand{\aap}{A\&A}
\newcommand{\mnras}{MNRAS}

\title[Period fluctuations in V1154~Cyg]{Period and light curve fluctuations of the {\it Kepler} Cepheid V1154~Cyg}

\author[A. Derekas et al.]{A. Derekas$^{1,2,3}$\thanks{E-mail:
derekas@konkoly.hu} \thanks{Hungarian E\"otv\"os Fellow and Magyary Zolt\'an Postdoctoral Research Fellow}, Gy. M. Szab\'o$^{1}$, L. Berdnikov$^{4,5}$, R. Szab\'o$^{1}$, R. Smolec$^{6}$,  L. L. Kiss$^{1,3}$, 
\newauthor L. Szabados$^{1}$, M. Chadid$^{7}$, N. R. Evans$^{8}$, K. Kinemuchi$^{9}$, J. M. Nemec$^{10,11}$, S. E. Seader$^{9}$,
\newauthor J. C. Smith$^{9}$, P. Tenenbaum$^{9}$\\  
\\$^1$Konkoly Observatory, Research Centre for Astronomy and Earth Sciences, Hungarian Academy of Sciences,
\\ H-1121 Budapest, Konkoly Thege Mikl\'os \'ut 15-17, Hungary
\\$^2$Department of Astronomy, E\"otv\"os University, Budapest, Hungary
\\$^3$Sydney Institute for Astronomy, School of Physics, University of Sydney, NSW 2006, Australia
\\$^4$Sternberg Astronomical Institute of Moscow University, Universitetskii pr. 13, Moscow, 119992 Russia
\\$^5$Isaac Newton Institute of Chile, Moscow Branch, Universitetskij Pr. 13, Moscow 119992, Russia 
\\$^6$Copernicus Astronomical Centre, ul. Bartycka 18, 00-716 Warszawa, Poland
\\$^7$Universit\'e Nice Sophia-Antipolis, Observatoire de la C\^ote d'Azur, UMR 7293, Parc Valrose, 06108 Nice Cedex 02, France
\\$^8$Smithsonian Astrophysical Observatory, MS 4, 60 Garden Street, Cambridge, MA 02138, USA
\\$^9$NASA Ames Research Center, MS 244-30, Moffett Field, CA 94035, USA
\\$^{10}$Department of Physics \& Astronomy, Camosun College, Victoria, British Columbia, V8P 5J2, Canada
\\$^{11}$International Statistics \& Research Corporation, PO Box 39, Brentwood Bay, British Columbia, V8M 1R3, Canada
}

\begin{document}

\date{Accepted ... Received ..; in original form ..}

%\pagerange{\pageref{firstpage}--\pageref{lastpage}} \pubyear{2002}

\maketitle

\begin{abstract}

We present a detailed period analysis of the bright Cepheid-type variable star V1154~Cygni ($V$=9.1 mag, $P$$\approx$4.9 d) based on almost 600 days of continuous observations by the {\it Kepler} space telescope. The data reveal significant cycle-to-cycle fluctuations in the pulsation period, indicating that classical Cepheids may not be as accurate astrophysical clocks as commonly believed: regardless of the specific points used to determine the $O-C$ values, the cycle lengths show a scatter of 0.015-0.02 days over the 120 cycles covered by the observations. A very slight correlation between the individual Fourier parameters and the $O-C$ values was found, suggesting that the $O-C$ variations might be due to the instability of the light curve shape. Random fluctuation tests revealed a linear trend up to a cycle difference 15, but for long term, the period remains around the mean value. We compare the measurements with simulated light curves that were constructed to mimic V1154~Cyg as a perfect pulsator modulated only by the light travel time effect caused by low-mass companions. We show that the observed period jitter in V1154~Cyg represents a serious limitation in the search for binary companions. While the {\it Kepler} data are accurate enough to allow the detection of planetary bodies in close orbits around a Cepheid, the astrophysical noise can easily hide the signal of the light-time effect.

\end{abstract}


\section{Introduction}

Cepheids are luminous, F and G type supergiant stars, exhibiting radial pulsations driven by the $\kappa$-mechanism with periods from a few days up to about 100 days. Owing to their high luminosity and well-defined period-luminosity relations, they are primary distance indicators. The pulsation period is one of the most important parameters of a Cepheid variable and it is assumed to be stable on short time-scales. However, on the evolutionary time-scales, Cepheid periods are subject to variations but these period changes become detectable only over several decades or even longer time-scales \citep{sza83,tur06}.

Observations of Cepheids with ground based instruments provide very poor light curve sampling, usually 1-2 data points per night which is mainly due to the long pulsation period of these variables. In addition, except whole Earth campaigns, observations usually cover a few weeks at most. The three phases of the Optical Gravitational Lensing Experiment (OGLE, \citet{sos08a,sos08b,sos10}) and MACHO \citep{alc99} projects produced years of observations of Cepheids in the Magellanic Clouds  but again, their light curves contain one point per night. While this is perfectly enough to characterize the general variability of the few thousand Cepheids, detailed insight into the light curve shape changes or short-term period fluctuations is prevented by the daily sampling. \citet{pol08} has succesfully analysed the period changes of Cepheids based on the OGLE and MACHO datasets and found variability on the scale of a few years. The near-continuous observations of {\it Kepler} give us a unique opportunity to study these variables in more detail than never before. 

Cepheids are often regarded as clockwork-precision, regular pulsators and with a few exceptions (V473 Lyr: \citet{bur82}, Polaris: \citet{tur05,bru08,spr08}) this is usually a valid approximation to the limits set by the sampling and precision of ground-based observations. State-of-the-art one-dimensional hydrocodes (Florida-Budapest: \citet{kol02}, Warsaw: \citet{smo08}) have been supporting this view, because after reaching their limit-cycle (in case of a single-mode pulsation) they are able to repeat pulsational cycles essentially forever. This of course assumes that no evolutionary or additional processes acting on longer time-scales are included in the set of equations solved numerically.

Photometric time series data obtained by space based telescopes have been available on a few Cepheids: \citet{spr08} analysed photometric data of Polaris observed with the SMEI instrument on board the Coriolis spacecraft; independently, \citet{bru08} also studied Polaris using the SMEI and WIRE star tracker photometry; \citet{ber10} studied the periodicity of eight Cepheids based on SMEI photometry; while the existing WIRE data on $\delta$~Cephei are currently being analysed \citep{bru07}. Although the precision of these data is superior to that of their ground based counterparts, the 0.0001~mag relative accuracy still hides the subtle effects to be discovered from the {\it Kepler} data.

In this paper we present the first evidence that the pulsation period of V1154~Cyg is jittering, i.e. it is not as stable as the present models predict. In Sect. 2 we briefly describe the data used, with details of pixel-level photometry of the saturated {\it Kepler} observations, the methods of the analysis and simulated study about the timing accuracy of the {\it Kepler} data. In Sect.\ 3, our findings on the period fluctuations in V1154~Cyg are described, while the implications are discussed in Sect.\ \ref{application}. We conclude with a short summary in Sect.\ \ref{summary}.

\section{Data analysis}

The data we use in this paper were provided by the {\it Kepler} space telescope. The telescope was launched in March 2009 and designed to detect transits of Earth-sized planets. A detailed technical description of the {\it Kepler Mission} can be found in \citet{koc10} and \citet{jen10a,jen10b}.

The {\it Kepler} space telescope observes 105 square degree area of the sky in the constellations Cygnus and Lyra, and has two observational modes, sampling data either in every 58.9~s (short-cadence whose characteristics was analysed by \citet{gil10}) or 29.4~min (long-cadence), providing near-continuous time series for hundreds of thousands of stars. The data are divided into quarters: the commissioning run, Q0 ($\sim$10~d), then Q1 ($\sim$34~d) followed by Q2, Q3, Q4, Q5, Q6, and Q7 ($\sim$90~d for each); when this paper was written $\sim$600 days of observations was available to us.

In the field of view of {\it Kepler}, there is only one genuine Cepheid variable so far \citep{sza11}, V1154~Cyg (KIC~7548061) which has been observed in long-cadence mode in all quarters and short-cadence mode in Q1, Q5, Q6. V1154~Cyg has a mean V brightness of $\sim$9.1~mag and period of $\sim$4.925~d. Previous observational studies of this Cepheid are listed and discussed by \citet{sza11}.

\subsection{Cepheid pixel photometry}

\begin{figure*}
\begin{center}
\includegraphics[width=12cm]{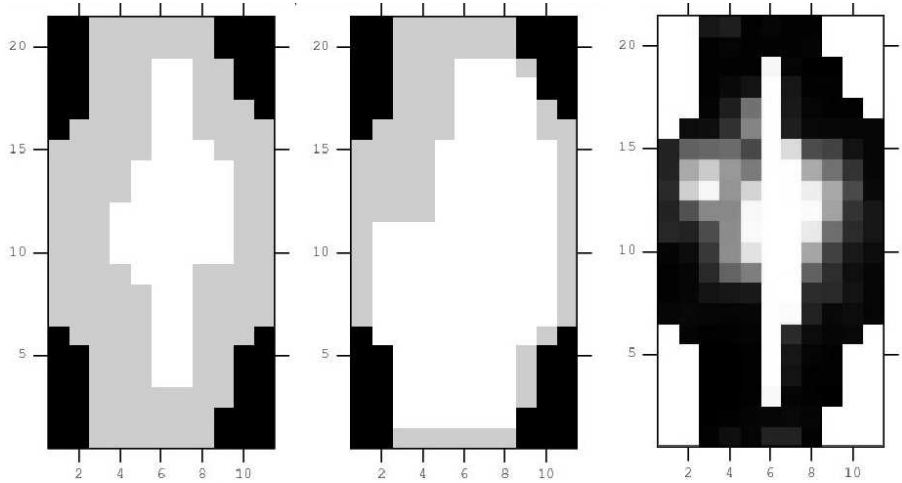}
\end{center}
\caption{\label{pixel} Left: original optimal aperture mask in Q2 (white), all the downloaded pixels (grey), pixels that were not downloaded (black). The axes label row and cloumn pixel numbers. Middle: our custom mask designed to retain all the flux. Right: Flux distribution around one of the maxima of V1154~Cyg in Q2 (greyscale). We deselected the pixels assigned to the near neighbour of V1154~Cyg.}
\end{figure*}

In \citet{sza11} we demonstrated that the apparent large amplitude variations (especially in Q2) are not intrinsic to the star, because some of the flux was lost from the assigned {\it optimal aperture} \citep{bry10} of V1154~Cygni. The usual output of the {\it Kepler} photometer is the integrated flux (light curve). However, to correct for this instrumental effect we have to rely on additional information to which we had no access at the time of writing the \citet{sza11} paper. 

Besides the extremely precise measurements of {\it Kepler} it is equally important that recently the individual pixel time-series data have been made available\footnote{MAST, http://archive.stsci.edu/kepler/} by the {\it Kepler} Team. In a number of cases the pixel data provided crucial  information (1) on the photocenter of the transit variations \citep{bat10} thereby confirming planetary candidates, (2) helped to devise a custom aperture  for RR Lyrae \citep{sza10}, which is too bright and is located very close to the edge of one of the CCDs, thereby providing a way to observe this important prototype Blazhko variable with {\it Kepler}, or (3) in a recent paper \citep{sau11} based on pixel data we were able to determine which component of a close visual common proper motion binary is transited by a 'hot' brown dwarf or planet.

We have downloaded all the long cadence target pixel files (Q0-Q7) of V1154~Cygni. We assigned a much larger aperture than the original optimal aperture to sum the pixels to ensure that we do not loose flux, see Fig.1. It is worth noting that V1154~Cygni is heavily saturated on the {\it Kepler} CCDs, so an elongated aperture is applied. It is also noticeable that the central (saturated) column is not contained in its entirety with the original optimal aperture. We took care to exclude additional stars captured in the downloaded pixels. Later it proved to be a wise decision, because by adding up its pixels the additional star centered at (3,13) turned out to be a previously unknown variable star showing frequency peaks in the [2,3] c/d frequency range.

The resulting light curve is very similar to the {\it Kepler} light curve, except that the spurious amplitude drop has disappeared. Although light curve sections of different Qs had to be shifted vertically because of the different sensitivities of the CCDs (occurring due to the quarterly roll of the space telescope), we did not have to adjust the amplitudes which is reassuring. The only exception is Q0 (commissioning phase containing only two pulsational cycles), where inspection of the downloaded pixels confirmed that some flux was lost in the maximum phases, because the flux spilled out of the downloaded pixels, and this affected the pulsation amplitude and the $O-C$ values as well. 

\subsection{The \boldmath{$O-C$} diagram}
\label{ocdia}

\begin{figure*}
\begin{center}
\includegraphics[width=17cm]{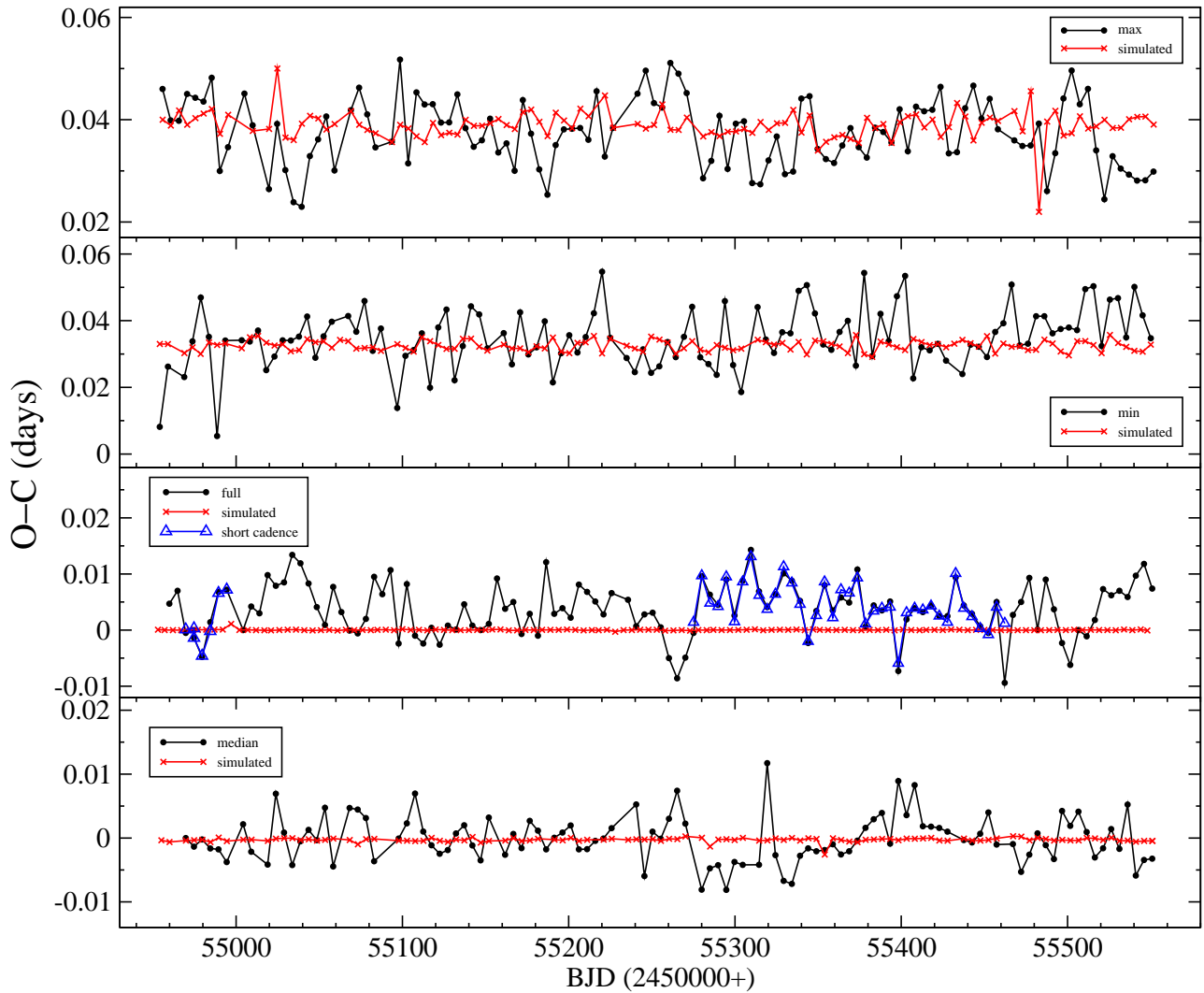}
\end{center}
\caption{\label{v1154cygoc} The $O-C$ diagram of V1154~Cyg calculated with 4 different methods (black dots) using the long cadence data, while blue triangles represent the $O-C$ diagram based on the short cadence data. The full description of the 4 methods is in Section\ \ref{ocdia}. Red crosses are the resulted $O-C$ diagrams of synthetic data, see in Sect.\ \ref{synt}.}
\end{figure*}

We studied the period stability of V1154~Cyg applying the classical $O-C$ diagram method \citep{ste05}. Thanks to the continuous observations of {\it Kepler}, we could determine the $O-C$ values for almost all cycles, except those with gaps. For a thorough analysis, we used 4 different methods to calculate the $O-C$ values that are described below.

The first and second methods were to determine the times of maxima and minima of the light curve by fitting tenth-order polynomials around the extrema. The calculated $O-C$ diagrams for the times of maxima and minima are shown in the top two panels of Fig.\ \ref{v1154cygoc} plotted with black dots. The average error of the $O-C$ values is $\pm$0.0021~d, which was determined by applying our method to a synthetic light curve (see Sect.\ \ref{synt}) generated using the Fourier parameters of V1154~Cyg and averaging the deviations between the analytically derived maxima/minima and the maxima/minima calculated by our method.

The third method was to measure phase shifts of each pulsational cycle and transform this to an $O-C$ diagram. This was done by fitting eighth-order Fourier-polynomial to the phase diagram of one pulsational cycle and used it as a master curve -- allowing only vertical and horizontal shifts -- to fit every other phase diagram. Combining the phase shifts and the period with the epoch, we calculated the $O-C$ values (with an average error of $\pm$0.0012~d which was adopted from the standard error of the fit of the master curve) which are plotted in the third top panel of Fig. \ref{v1154cygoc} with black dots.

We also determined the moments of the median brightness (mid value of the adjacent minimum and maximum brightness) on the ascending part of the light curve. Their $O-C$ diagram are plotted with black dots in the bottom panel of Fig.\ \ref{v1154cygoc}. Assuming a 0.0001 relative uncertainty in the value of both the minimum and maximum brightnesses of a given epoch number, the moment of the median brightness can be determined with an unprecedented precision of 0.0001 d, owing to the steepest brightness variation during this particular phase of the pulsation cycle. The significantly smaller uncertainty of the median $O-C$ values testifies that, for photometric observations of low amplitude Cepheids, it is the point of the median brightness on the ascending branch that is the preferred light curve feature for following the behaviour of the pulsation period. 

For the calculation of the $O-C$ diagrams ({\it O} stands for the observed, {\it C} stands for calculated times of extrema), we used the following ephemeris: \begin{equation}C=HJD~2454969.70432+4.925454 \cdot E, \end{equation} where {\it C} is the predicted time of the maxima (i.e. C), while {\it E} refers to the cycle number. The $O-C$ values show a scatter of about $\pm$0.015 days ($\cong$20 min) for the full and the median and about $\pm$0.02 days ($\cong$30 min) for the maxima and the minima. The scatter of the $O-C$ diagram is far larger than we expect from a Cepheid variable (see Sect.\ \ref{results} and Table\ \ref{maxtimes}).

We also calculated the $O-C$ diagram using the available Q1, Q5 and Q6 short cadence data by applying the third method. The results are plotted with blue triangles in the third panel of Fig.\ \ref{v1154cygoc}. There is no significant difference between the resulted $O-C$ diagrams whether it is based on the long cadence and the short cadence data, giving a supporting evidence that the detected period jitter of V1154~Cyg does not originate from the data sampling. As it is expected, however, the average error of the short cadence $O-C$ values is significantly smaller than the long cadence one: $\pm$0.0001~d.

\subsection{\label{synt} Timing accuracy of Kepler data}

To check the reliability of the detected $O-C$ variations with various methods, we designed a numerical experiment, based on synthetic data of Kepler quality, describing a hypothetic stable light variation (no changes in light curve or pulsation phase). The model was a 6th order Fourier-polynomial, fitted to the entire light curve. In such a way, the model described the average light curve shape of V1154 Cyg.

To produce artificial data, bootstrap of {\it Kepler} noise was added to the model with null signal. The noise was taken from the time series of V1154 Cyg: the residuals were cycle-by-cycle fitted by a Fourier-polynomial, and the set of residuals were sampled with a replacement. Then artificial data were sent into the same algorithms as in the case of the observations. Since the measured distribution of $O-C$ regards to a null-signal event, it exactly reflects the accuracy of timing measurement in {\it Kepler} time series of V1154 Cyg.

$O-C$ of artificial data were non-Gaussian, they exhibited significantly heavier wings than a normal distribution. This is because 2--4 points are severe outliers, and they reflect probably the effect of uneven sampling (i.e. gaps in data). However, the method of the median brightness and the fit of the full light curve still results in standard deviations of 0.00018 and 0.00021 days, respectively, very small values compared to the measured effect.

The methods of maximum and minimum brightness exhibit a significantly larger scatter of $\approx0.01$ days, but after neglecting the four most distant outliers, the scatter decreases to 0.002 and 0.003 days, respectively. This scatter is larger than in the case of the median brightness or fitting the full light curve, compared to measuring at the maximum or minimum of the light curve when the brightness variation is slow. However, these values are still many factors smaller than the detected signal. In summary, we conclude that the variation of $O-C$ diagrams cannot be interpreted with numerical fluctuations, and the measured $O-C$ patterns are real.

\begin{figure}
\begin{center}
\includegraphics[bb=64 55 415 300,width=\columnwidth]{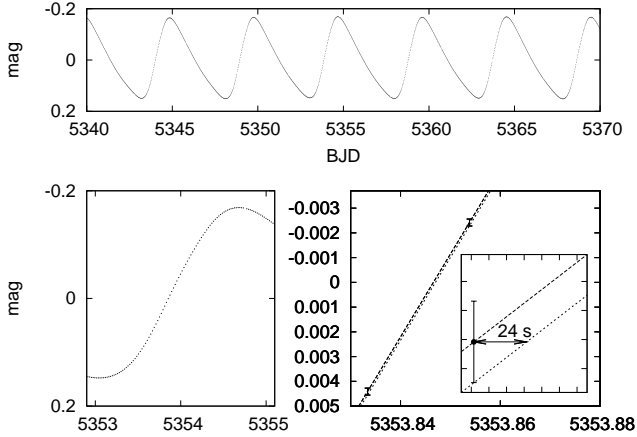}
\end{center}
\caption{\label{median} Demonstration of the accuracy of timing the median brightness. The upper panel shows six cycles of the light curve. The lower left panel is a magnification of one individual ascending branch, and the lower right panel is a zoom into the region between two consecutive data points that bracket the median brightness. The dashed line in the small inset corresponds to the best-fit linear function, while the dotted line is the lowest limit that still goes through both errorbars. The time-lag between the two lines is an estimate of the accuracy, in this case 24 seconds. }
\end{figure}

In Fig.\ \ref{median}, we explain the timing method using median brightness. The upper panel shows a segment of the light curve, while the lower panels zoom in two regions of the same light curve. . The error bar of the individual data points (0.1 mmag) can still hardly be seen in the lower right panel, which spans 8.5 mmag in the  vertical axis. Here we compare the best-fit linear function (dashed line) through the two consecutive data points that bracket the median brightness. The dotted line corresponds to the limiting case where the linear function is shifted vertically exactly to the lowest limits of the error bars, keeping the fit still consistent with the data. The time lag between the dashed and the dotted lines is 24 seconds; solutions with a larger time lag would misfit at least one of the data points, hence we take this measurement as an estimate of the accuracy. 

The accuracy in timing of photometric minima and maxima is worse because the light curve is flat at the extrema, and the error bars enable somewhat larger time uncertainties.

\subsection{Fourier parameters}

We studied the change of the light curve shape as a possible cause of the scatter in the $O-C$ diagram by examining the temporal variation of the Fourier parameters of the light curve. For this, we fitted eighth-order Fourier polynomial at the primary frequency and its harmonics: \begin{equation} m=A_{0}+\sum_{i=1}^{8} A_{i} \cdot \sin\left( 2 \pi ift + \phi_{i} \right), \end{equation} where {\it m} is the magnitude, {\it A} is the amplitude, {\it f} is the frequency, {\it t} is the time of the observation, $\phi$ is the phase and index {\it i} runs from 1 to 8. Then we characterised the light curve shapes with the Fourier parameters \citep{sim82}, of which we show particular results for $R_{21}=A_{2}/A_{1},\ R_{31}=A_{3}/A_{1} \ {\rm as \ well \ as}\ \phi_{21}=\phi_{2}- 2\phi_{1}\ {\rm and} \ \phi_{31}=\phi_{3}-3\phi_{1}$,

\begin{figure*}
\begin{center}
\includegraphics[width=17cm]{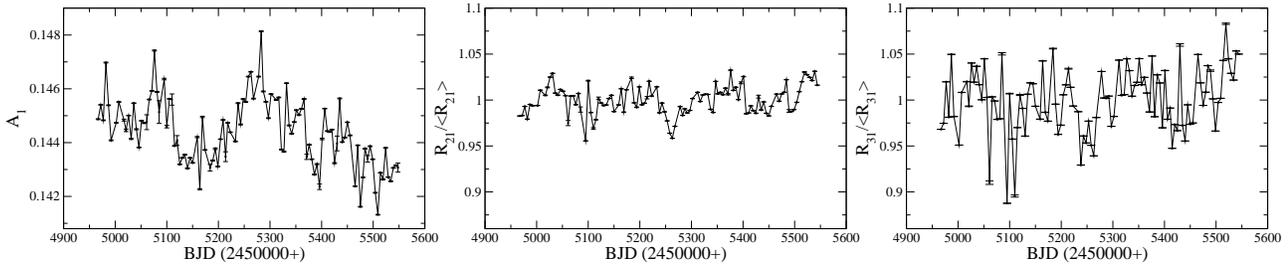}
\end{center}
\caption{\label{fourpar} From left to right: the amplitude ($A_{1}$) change and the relative variations of $R_{21}$ and $R_{31}$ with average errors of $\pm$ 0.000042, $\pm$0.0001, $\pm$0.0012, respectively.}
\end{figure*}

\begin{figure}
\begin{center}
\includegraphics[width=7.5cm]{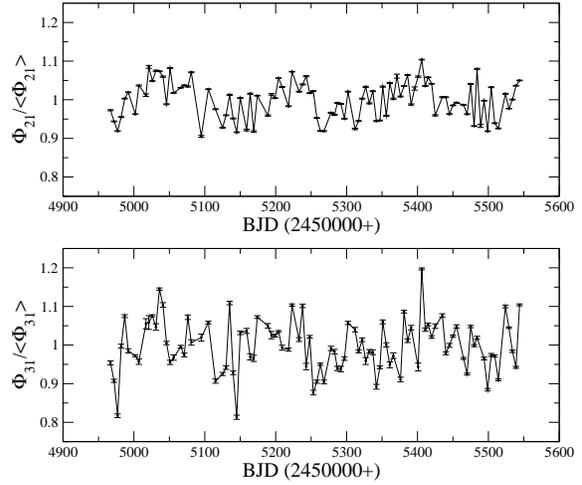}
\end{center}
\caption{\label{fourpar1} The relative phase variations of $\phi_{21}$ (top) and $\phi_{31}$ (bottom), with average errors of $\pm$0.0013, $\pm$0.0051, respectively.}
\end{figure}

Fig.\ \ref{fourpar} shows the amplitude change and the relative variations of $R_{21}$ and $R_{31}$, while Fig.\ \ref{fourpar1} shows the relative variation of $\phi_{21}$ and $\phi_{31}$. All of these parameters show significant cycle-to-cycle variations. Detailed study of these variations in the $O-C$ and the Fourier parameters is described in the next section.

\section{Results}
\label{results}

\subsection{Correlations between \boldmath{$O-C$} datasets}

Different determinations of $O-C$ points show loose correlations, no correlation or even anticorrelations with each other. Results of a Spearman's rank correlation test \citep{lup93} is summarized in Table 1; for each compared dataset pairs, the top row shows Spearman's $\rho$ and the $p$-value is given below. This test is like a generalization of the correlation coefficient: the value of $\rho$ measures how well the dependence between two variables can be described with a general monotonic function. The value of $\rho$ is 1 or $-$1 if there is a perfect fit with an increasing/decreasing function, respectively; values in between express the relative weight of residuals which cannot be explained by a monotonic law. The value of $p$ shows how probable is that the suspected connection between the variables is due to a false recognition of simple numerical fluctuations. {\it Id est,} an intermediate value of $\rho$ still safely expresses the slight monotonic connection between the variables if $p$ is very small. The significant correlations are as follows:
$$(O-C)_{\rm Max}=0.7(2)\times (O-C)_{\rm Med} + 1148.6(9) $$
$$(O-C)_{\rm Full}=-0.6(1)\times (O-C)_{\rm Med} + 5(5)$$
$$(O-C)_{\rm Full}=-0.52(4)\times (O-C)_{\rm Max} + 600(50)$$
$$(O-C)_{\rm Max}=-0.11(7)\times (O-C)_{\rm Min} + 1010(90)$$ 

\noindent (errors of the last digits are given in parentheses, absolute terms are in seconds).

The most interesting is the general lack of strong dependences between the differently defined $O-C$ values derived for the same pulsation cycles. In particular, the minimum is practically not connected to any other $O-C$ measures; while $O-C$ of the maxima shows relatively good (but still somewhat stochastic) dependence on the $O-C$ of the whole cycle. Another remarkable feature is the anticorrelation between the $O-C$ of the whole cycle and that of the median brightness or the maximum.

This behaviour suggests two conclusions. At first, the scatter in $O-C$ values is somehow reproducible, at least certain kinds of $O-C$ values can be predicted to some accuracy if one knows the value of another $O-C$. Therefore, the scatter we observed in any $O-C$ is not due to some unrecognized stochastic error, e.g. coming from the sampling and/or the evaluation of data. Perhaps the phase of the pulsation is changing, or there are slight local irregularities in the light curve. But the second finding is that the process which acts on the $O-C$ values changes rapidly, this is why there are no strict correlations between the different determinations of $O-C$. These together suggest that the cause of the observed behaviour of $O-C$ data is the irregular local fluctuations of the light curve shape. We  shall explore this hypothesis in the next subsections.

\begin{table}
\caption{Results of a Spearman's rank correlation test of different $O-C$ datasets. The sign $ns$ means no correlation with a confidence exceeding 85\%{}.}
\begin{tabular}{lccc}
\hline
 & Min & Max & Full\\
\hline
Med &  $ns$ & 0.42 & $-$0.51\\
   &     & $p=5\cdot 10^{-6}$ & $p=2\cdot 10^{-8}$ \\
\hline
Min &  ---& $-$0.14 & $ns$ \\
   &     & $p=0.15$ \\
\hline
Max & --- & ---&$-$0.73\\
   &     &    &$p=2\cdot 10^{-16}$  \\
\hline
\end{tabular}
\end{table}

\subsection{Fourier parameters explain \boldmath{$O-C$} points}

Local variations of the light curve can be described by higher-order Fourier-coefficients. Therefore one expects if changes in the light curve shape were the primary reason for $O-C$ variations, then there would be connections between the Fourier parameters and the $O-C$ values. If linear models are taken into account, only approximate fits are expected because the light curve shape depends on the Fourier parameters in a very complex and non-linear way. This was really observed: we generally found very slight correlations between the individual Fourier parameters and the $O-C$ values.

Another possibility is a multilinear fit of the $O-C$ points using the Fourier parameters. Taking the first 8 amplitude and phase coefficients, and additionally, the height of the minimum and the maximum of the individual cycles, we found that the times of the mean amplitude can really be predicted with 90\%{} accuracy via a linear -- and therefore rather heuristic -- model (i.e. the scatter decreases by 10\%{} meanwhile the degree of freedom decreases by only 17\%{}, from 126 to 108). The accuracy was less impressive in the other cases, e.g. it was only 30\%{} when fitting the times of minima (i.e., the residuals decreased only by 30\%{} after the fit).

The probable interpretation is that the $O-C$ variations are really due to the instability of the light curve shape, because the light curve shape alone is the best predictor of $O-C$ values. The existence of less impressive fits does not contradict much this interpretation, because the light curve shape is a non-linear function of our parameter space, and it has been foreseen that the linear fit will not always work. In fact, good fits can be expected only in the case when the parameters are combined by fortune in such way that the local light curve irregularities can be effectively described with a multilinear approximation; as it happened for the times of median brightness.

\subsection{Eddington-Plakidis test}

The $O-C$ residuals (Fig.\ \ref{v1154cygoc}) were analyzed for the presence of random fluctuations in the pulsation period using the method described by \citet{edd29}.

\citet{edd29} made a formalism to test for random, cycle-to-cycle fluctuations in period. They proposed that the average value $ \langle u(x) \rangle $ is given as\begin{equation} \langle u(x) \rangle ^2 = 2 \alpha^2 + x \varepsilon^2 ,\end{equation} where $ \langle u(x) \rangle $ is the mean absolute difference between the $O-C$'s which are $x$ cycles apart, $\alpha$ is the mean observational error in measuring the time of maximum or minimum light and $\varepsilon$ is the mean fluctuation in period from one cycles to the next \citep{per00}. These errors are each assumed to be accidental and uncorrelated. Hence a plot of $\langle u(x)\rangle ^2$ against $x$ should be a straight line with slope $\varepsilon^2$ and intercept $2 \alpha^2$.

\begin{figure}
\begin{center}
\includegraphics[width=8cm]{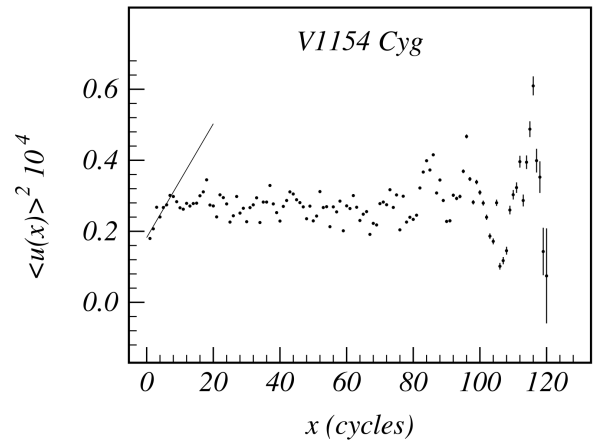}   
\end{center}
\caption{\label{ep} The result of the Eddington-Plakidis test. We fit a linear trend up to x=15, which indicates fluctuations in the period or light curve shape. See text for the full description of this effect.}
\end{figure}

In practice, this means that if there is a linear trend in the plot of $ \langle u(x) \rangle ^2$ at low $x$ (from several to several tens cycles), then it is assumed that random period fluctuations are present in the pulsations of a given star and this segment of the plot is used for a linear fit. In other words, all wave-like fluctuations in the $O-C$ residuals (after the removal of all trends) are assumed to be attributed to random fluctuations in the pulsation period or in the light curve shape.

As it is shown in Fig.\ \ref{ep}, a linear trend is revealed up to a cycle difference x$\sim$15. The mean random fluctuation parameters are summarised in Table\ \ref{eptable} for the four different methods.

\begin{table}   
\begin{center}
 \caption{\label{eptable} Coefficient solutions of the Eddington-Plakidis test for four different methods.}  
\label{maxtimes}  
\begin{tabular}{rcccc}  
\hline   
Method & $\epsilon$ & $\epsilon_{{\rm err}}$ & $\alpha$ & $\alpha_{{\rm err}}$\\ 
\hline  
Full & 0.0013 & 0.00054 & 0.0030 & 0.00085 \\
Med & 0.0008 & 0.00042 & 0.0023 & 0.00066\\
Min & 0.0015 & 0.00058 & 0.0063 & 0.00141\\
Max & 0.0015 & 0.00048 &  0.0043 &0.00105  \\
\hline   
\end{tabular} 
\end{center}  
\end{table}

These results lead us to the conclusion that in short terms (up to a cycle difference x$\sim$15), the period and/or the light curve shape changes randomly. However, for long term, there is an internal clock in V1154~Cyg which keeps the period around the mean value. This feature is also supported by the fact that the $O-C$ diagram of V1154~Cygni covering 15000 days indicates a constant mean pulsation period (see Fig.~13 in \citet{sza11}).

A similar Eddington-Plakidis test was performed for SMEI photometric data of 8 bright Cepheids by \citet{ber10}. He found random period fluctuations, in some cases superimposed on period changes due to stellar evolution, but the period and its changes could not be followed on a time-scale of individual pulsation cycles unlike this study of V1154~Cyg.

\section{Implications of the results}
\label{application}

If the results presented here for V1154~Cyg are common among Cepheids then the consequences might affect other areas of research.

\subsection{Cepheids models}

Numerical hydrodynamical models do not show such hectic period variations from cycle-to-cycle. This is primarily because these codes contain simplified description of turbulent convection (Florida-Budapest code \citet{kol02}, Warsaw code \citet{smo08}), or before the early '90s contained radiative energy transfer only. Using this type of physics the pulsation usually reaches its limit cycle (or alternatively double-mode pulsation), then stays in this state essentially forever, because these codes are very good at conserving energy and momentum. Hence, the pulsation period is practically constant. Any deviation from this precise, regular machinery is due to choosing imperfect numerical scheme or coding errors. The above mentioned codes are free from these problems.

Present-day models are not capable of dealing with complex, magneto-hydrodynamic interaction \citep{sto09} that could cause regular or irregular modulations of the stellar structure (hence in the period), because of the enormous difficulties of implementing and computing 3D MHD for full stellar envelopes. However, \citet{buc93} developed a method based on the amplitude equation formalism to describe convective and turbulent random fluctuations and hence stochastic interaction between convection and pulsation. This mechanism would cause fluctuations in the period, but would not deform the shape of the pulsation.

Deviations from this precise clockwork mechanism do appear in certain high-luminosity models \citep{buc87,kov88,mos90,mos91,buc92}, namely, they can show period doubling or a series of period doubling bifurcations en route to chaos. These types of dynamical phenomena also cannot be responsible for the observed period variations.

Evolutionary changes are usually happening on a much longer time-scale and in a regular manner, in addition the physics driving the evolution is usually missing from these models designed to follow only the pulsation.

Clearly, a more precise (preferably 3-dimension) description of turbulent convection would be essential to model the observed large cycle-to-cycle period `jitter'.

\subsection{Light curve distortion}

We can speculate that a stable global pulsation can appear as 'jittered' or distorted provided the stellar atmosphere is not quasi-static or spherically symmetric due to large-scale turbulent cells for example. This effect provides no feedback to the global pulsation, but perturbs the observed light curve. Changes causing local perturbations of the optical depth may also induce apparent phase disturbances. This effect may be present in all cool giant variable stars although with different magnitude. However, the detailed modeling of these mechanisms are beyond the scope of the present paper.

\subsection{Searching for companions}

The usually assumed stability of the pulsation period is a fundamental ingredient in finding low mass or substellar companions around Cepheids. In light of the period jittering found in V1154~Cyg, it is worth checking what kind of sensitivity can be expected for a perfect pulsator observed by {\it Kepler}. For this, we have calculated several simulations as follows. First we took a short subset of the long cadence light curve covering one pulsation cycle to fit a 4th order Fourier polynomial that represented the ideal light curve shape of the star. Then we took each time stamp of the combined Q0-Q7 long cadence light curve to calculate the analytic template, which was phase modulated by a sinusoidal term: the period of the modulation was set to 200 days, while the amplitude was set to representative values between 0.1 s and 60 s. The flux errors reported by the {\it Kepler} pipeline were used to add a random error to each point of the simulation. This way the properties of the random errors matched those of the original data. The simulated light curves were then analysed with the fourth method used in the $O-C$ analysis (see Sect.\ 2.2), namely with the local phase determination using a master curve represented 
by a Fourier polynomial.  

\begin{figure}
\begin{center}
\includegraphics[width=8.5cm]{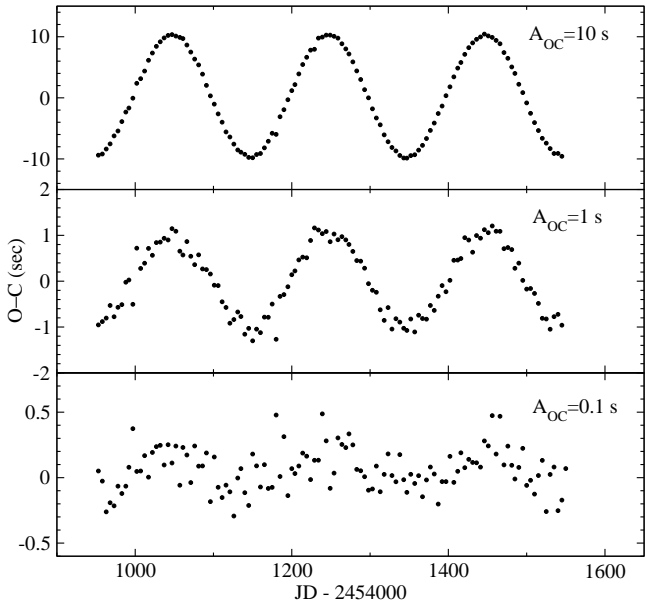}   
\end{center}
\caption{\label{models} The $O-C$ diagram of three model light curves with sinusoidal phase modulations (modulation semi-amplitudes shown in the upper right corner in each panel). See text for details.}
\end{figure}

In Fig.\ \ref{models} we show the results for three simulations. The amplitudes of the phase modulations were 10 s, 1 s and 0.1 s, which can be connected to the mass ratio of the companion and the Cepheid using the third Kepler-law and assuming a mass for the Cepheid. Ignoring the $\sin i$ ambiguity of the unknown inclination angle (i.e. assuming an edge-on view of the system), the semi-amplitude of the $O-C$ diagram depends on the Cepheid mass and the mass ratio as follows: 

\begin{equation}
A_{OC}=\frac{1}{c}\left(\frac{G}{4\pi^{2}}\right)^{1/3}\frac{M_{1}^{1/3}\left(1+\frac{M_{2}}{M_{1}}\right)^{1/3}}{\left(1+\frac{M_{1}}{M_{2}}\right)}P_{\rm orb}^{2/3}
\end{equation}

\noindent where $M_{1}$ and $M_{2}$ are the masses of the Cepheid and its companion, respectively, $P_{\rm orb}$ is the orbital period, $G$ is the gravitational constant, $c$ is the speed of light. For a 5 M$_{\odot}$ Cepheid and a 200 days orbit, the given three $O-C$ amplitudes correspond to a companion mass of about 50 M$_{\rm Jup}$, 5 M$_{\rm Jup}$ and 0.5 M$_{\rm Jup}$.

The three simulated O$-$C diagrams in Fig.\ \ref{models} clearly show that {\it Kepler}'s precision could be enough to detect planet-sized companions on short-period orbits around Cepheids, if the stars were perfect clocks. However, the astrophysical noise in V1154~Cyg is way too high to allow that kind of detections of substellar companions. 

\section{Summary}
\label{summary}

The period variations for the Cepheid V1154~Cyg have been examined based on 600 days of photometric observations with the {\it Kepler} space telescope. In order to analyse the best quality light curve we performed pixel photometry on the available individual pixel time-series data. We calculated $O-C$ values for all cycles using 4 different methods: times of maxima and minima, median brightness and phase shifts of all pulsational cycles. We also determined the Fourier parameters of each cycle.

We found very slight correlation between the individual Fourier parameters and the $O-C$ values, suggesting that the $O-C$ variations might be due to the instability of the light curve shape. Random fluctuation test revealed linear trend up to a cycle difference 15 but for long term the period remains around the mean value.

Finally, we showed that if the period jitter is common among other Cepheids then it needs to be taken into account in the hydrodinamic models and sets a limit to detect substellar companions around Cepheids.

Subtle variations in the pulsation period and shape of the light curve may be present in the pulsation of other Cepheids. Indirect evidence supporting the behaviour mentioned here was given by \citet{kla09} who found that the photometric phase curves of some small amplitude Cepheids in our Galaxy show a wider scatter than expected from the measurement errors even if data are folded on the correct value of the pulsation period.

\section*{Acknowledgments} 

This project has been supported by the Hungarian OTKA Grants K76816, K83790 and MB08C 81013, ESA PECS C98090 and the ``Lend\"ulet-2009'' Young Researchers Program of the Hungarian Academy of Sciences. The research leading to these results has received funding from the European Community's Seventh Framework Programme (FP7/2007-2013) under grant agreement no. 269194 (IRSES/ASK). AD gratefully acknowledges financial support from the Magyary Zolt\'an Public Foundation. AD was supported by the Hungarian E\"otv\"os fellowship. AD, GyMSz and RSz has been supported by the J\'anos Bolyai Research Scholarship of the Hungarian Academy of Sciences. Funding for this Discovery mission is provided by NASA's Science Mission Directorate. Fruitful discussions with Zolt\'an Koll\'ath are gratefully acknowledged.

\end{document}